\documentclass{aastex}
\usepackage{spr-astr-addons}
\usepackage{url}
\usepackage{color}
\RequirePackage{color}

\begin{document}

\title{Establishing a particle distribution for multi-wavelength emission from BL Lac objects}
\shorttitle{Establishing a particle distribution}
\shortauthors{Xie et al.}

\author{X.X. Xie\altaffilmark{1}} \and \author{K.R. Zhu\altaffilmark{1}}
\and
\author{S.J. Kang\altaffilmark{2}}
\author{Y.G. Zheng\altaffilmark{1}}
\affil{ynzyg@ynu.edu.cn}

\altaffiltext{1}{Department of Physics, Yunnan Normal University, Kunming, Yunnan, 650092, China}
\altaffiltext{2}{School of Physics and Electrical Engineering, Liupanshui Normal University, Liupanshui, Guizhou, 553004,  China}

\begin{abstract}
Electrons are accelerated at the shock wave diffuse and advect outward, and subsequently drift away into the emitting region of the jet that is located in the downstream flow from the plane shock. The current work considers the acceleration of the electrons in the shock front. Assuming a proper boundary condition at the interface between the shock and the downstream zones, a novel particle distribution in the downstream flow is proposed in this work to reproduce the broadband spectral energy distribution of BL Lac objects. We find that (1) we can obtain the particle distribution downstream of the shock wave in four cases; (2) electrons with higher energy ($\gamma>\gamma_{0}$) dominate the emission spectrum; (3) the distinctly important physical parameters assumed in our model can reasonably reproduce the multi-wavelength spectrum of the high-synchrotron-peaked BL Lac object Markarian 421 (Mrk 421).
\end{abstract}



\section{Introduction}

BL Lac objects are Active Galactic Nuclei (AGNs) characterized by a polarised and highly variable nonthermal continuum emission that extends from the radio to the TeV $\gamma$-ray bands. The multi-wavelength spectral energy distributions (SEDs) of these objects have two bumps shape. It is well known that the low-energy peak occurs at frequencies ranging from the infrared to the X-ray band, and the high-energy peak occurs at frequencies ranging from the MeV to TeV band \citep{{1981ApJ...243..700K,1998MNRAS.299..433F,2001A&A...371..512C}}.
Both the original lepton and the hadron models are the currently accepted mechanisms for the emission energy spectra of BL Lac objects \citep{{1985ApJ...298..114M,1996ApJ...461..657B,aleksic2012discovery}}. In the original lepton scenario, the low-energy peak mainly comes from the synchrotron emission of ultra-relativistic electrons in the jet with a small angle to the observer's line of sight, and the high-energy peak comes from the inverse Compton scattering of ultra-relativistic electrons \citep{{1993A&A...269...67M,1994ApJ...421..153S,2003ApJ...586...79A}}. Due to the different sources of the seed photons of the inverse Compton scattering, the lepton model can be divided into the External Compton (EC) model and the Synchrotron Self-Compton (SSC) model \citep{1992ApJ...397L...5M,1996MNRAS.280...67G,1996ApJ...463..555I,2008ApJ...686..181F,2016MNRAS.461.1862K,2016MNRAS.463.4481Z,2017ApJS..228....1Z}. In the original hadron scenario, it is considered that the high-energy emission comes from the cascade or synchrotron emission process of ultra-relativistic properties \citep{1999ApJ...510..188B,2014A&A...571A..83P}.
The method of simulating the SEDs is an excellent tool for studying jet physics \citep{{2009MNRAS.397..985G,2016arXiv160205965F,2017ApJ...842..129C}}.
It can be used to diagnose particle acceleration in blazars because the accelerated non-thermal particles in the jet  can detect broad-band continuous emissions \citep{celotti2008the,2010MNRAS.402..497G,2014ApJ...782...82D}. In this context, the potential particle distributions can be estimated from the observed multi-wavelength SEDs, and the particle acceleration mechanism can be constrained by the shape of those particle distributions \citep{{2016ApJ...819..156B}}. However, inversion of jet properties from the SEDs of blazars is a very tricky problem, due to observation limitations, we cannot directly obtain the picture of blazars.

The observed SEDs of blazars can be reproduced through a model that assumes a non-thermal relativistic electron energy distribution (EED). There are several kinds of steady-state electron distribution, including a simple power
law, broken power law \citep{{Albert2007Variable,2013ApJ...764..113Z}}, double broken power law \citep{{abdo2011insights}}, logparabolic, power law
with an exponential high-energy cutoff, and power law at low energies with a log-parabolic high-energy branch \citep{{2006A&A...453...47K,2009A&A...501..879T}}.

Many specific models were proposed to study jet emissions. The numerous previous modelling efforts for steady-state SEDs, used to describe blazars, assumed that some unspecified mechanism produced a particle distribution and then injected it into the emission region \citep{2019ApJ...873....7Z}. The required distribution of emitted particles was established by a variety of mechanisms, including first-order and second-order Fermi accelerations. The first-order acceleration was due to multiple shock waves crossing \citep{{1964ApJ...140.1013W,1978ApJ...221L..29B}}, where particle acceleration was accompanied by the acquisition of large amounts of energy, resulting in high-energy power-law emission. The second-order Fermi acceleration \citep{{1968JETP...26..821T,1971Ap&SS..12..302K}} was due to the interaction of magnetohydrodynamic (MHD) waves with random magnetic fields. It was used to describe the random process of particle energy diffusion and is an efficient acceleration mechanism. Besides, it also included the electrostatic acceleration due to magnetic reconnection.

Relativistic electrons are generally assumed to be injected into a shock front that passes through the emission region, starting close to the central object \citep{1998A&A...333..452K}. Then the energy spectrum forms a double power law distribution due to rapid radiation cooling and escape.
In the case of shock acceleration \citep{{1996ApJ...456..106D,1994ApJ...421..550D,2019ApJ...872...10X}},
the multi-wavelength spectral energy distribution is calculated by solving the electron transport equations in consideration of escape, synchronous acceleration and IC cooling \citep{{2016ApJ...819..156B}}.
Numerical (e.g., \citealt{2009A&A...501..879T,2018MNRAS.478.3855Z}) and analytical  (e.g., \citealt{2008ApJ...681.1725S,2009A&A...501..879T,2014ApJ...791...21F,2018ApJ...853....6L})
methods were used to solve the distribution of particles under certain hypothetical conditions.

Based on previous research results, it is theoretically possible to generate an SED by constructing particle transport equations and obtaining their solutions for blazars. In order to study the effect of particle distribution on the  photon spectrum for blazars, we build on the work of \cite{2019ApJ...873....7Z} and further investigate the spectrum of particles.
The plasma arriving the shock wave downstream moves relative to the observer. In order to gain insight into particle distribution and blazars, the present paper establishes a new  particle distribution in the downstream.
We assume that there exists a spherical blob in the jet, which is filled with extremely relativistic electrons and is evenly distributed isotropically in the emission region. The main purpose of this paper is to prove that the established particle distribution can reproduce multi-wavelength spectra under reasonable physical parameter assumptions.

The present paper is organized as follows. In Section 2, assuming a proper boundary condition at the interface between the shock zone and the downstream zone, we establish a particle distribution in the downstream flow. In Section 3, we use the established electron spectra to derive the calculation of the theoretical
photon spectrum in the frame of an SSC model. In Section 4, we apply the model to the quiescent-state emission from BL Lac object Mrk 421; discussions are given in Section 5. Throughout the paper, we assume the Hubble constant H$_0$=75 km{$\rm~ s^{-1} \rm~Mpc^{-1}$}, the dimensionless numbers for the energy density of matter $\Omega_M$=0.27, the radiation energy density $\Omega_r$=0, and dimensionless cosmological constant $\Omega_{\Lambda}$=0.73 in this paper.

\section{Establishing a particle distribution}

We consider that the jet travels along the cylinder at the shock front, with a certain velocity ratio of upstream and downstream. Electrons accelerate at the shock front and then drift downstream. This situation can be viewed as having two spatial regions: one in which the particles are constantly accelerating around the shock wave upstream, and the other in the downstream region in which the particles release a large amount of energy \citep{{1992ApJ...396L..39B}}.
In each region, the scattering is assumed to maintain a nearly isotropic distribution of particles \citep{{1998A&A...333..452K}}.
We adopt an accelerated steady state particle spectrum \citep{2019ApJ...873....7Z} as follows:
\begin{eqnarray}
N_{0}(\gamma, \gamma_{0})&=&\frac{\dot{N}_{0}m_{e}c}{4D_{0}\mu}
\begin{cases}
(\frac{\gamma}{\gamma_{0}})^{\alpha_{1}}, &\text{$\gamma_{min}\leq\gamma\leq\gamma_{0}$},\\
(\frac{\gamma}{\gamma_{0}})^{\alpha_{2}}, &\text{$\gamma_{0}<\gamma<\gamma_{max}$}.
\end{cases}
\label{Eq:1}
\end{eqnarray}
Where $\dot{N}_{0}$ is  the rate of continuous injection of the particle, $D_{0}$ is the momentum diffusion rate constant, $\gamma_{0}$ is the characteristic  Lorentz factor for the electrons injected into the shock upstream, and the injected particles have a mono-energetic distribution. $\gamma_{min}$ and $\gamma_{max}$ are the minimum and maximum Lorentz factors respectively. It is worth mentioning that $\alpha_{1}$ and $\alpha_{2}$  are electron spectral indices formed by the accelerated particles of the shock upstream. They are two roots that can be obtained by calculating the power-law solution of Green's function using the characteristic polynomial $\alpha^{2}-(2+\hat{A})\alpha-\hat{C}=0$ \citep{{2019ApJ...873....7Z}}. We can obtain the corresponding solution
\begin{equation}
\alpha_{1}=\frac{2+\hat{A}+\sqrt{(2+\hat{A})^{2}+4\hat{C}}}{2}\,,
\label{Eq:2}
\end{equation}
here, the  index $\alpha_{1}$ applies at low energies ($\gamma\leq\gamma_{0}$), and
\begin{equation}
\alpha_{2}=\frac{2+\hat{A}-\sqrt{(2+\hat{A})^{2}+4\hat{C}}}{2}\,,
\label{Eq:3}
\end{equation}
the index $\alpha_{2}$ applies at high energies ($\gamma>\gamma_{0}$).
$\hat{A}$ and $\hat{C}$ are dimensionless constants, previous efforts defined $\hat{A}$ as the ratio between the first-order momentum gain rate constant $\hat{A}_{0}$ and $D_{0}$, and $\hat{C}$ as the ratio between the shock regulated escape rate constant $\hat{C}_{0}$ and $D_{0}$ \citep{{2016ApJ...833..157K,2019ApJ...873....7Z}}, as we will explain in Section 4.

The particle spectrum $N_{0}(\gamma, \gamma_{0})$ is the power-law distribution formed by the injection and acceleration of low-energy particles in the upstream of the shock, which then drift downstream flow from the shock and lose energy. For power-law solutions, it is generally presumed that the energy losses are unimportant in the vicinity of the shock. It is believed that the efficiencies of cooling and $Bohm$ diffusive escape for the particles with lower energy are less than these for the particles with higher energy. Our model assumes that low-energy particles are injected in the upstream of the shock wave, the injected characteristic Lorentz factor $\gamma_{0}$ is much smaller than the equilibrium Lorentz factor in the system. Therefore, the evolution of particles in the upstream of the shock wave is dominated by the process of gain the energy. In this scenario, we can neglect the cooling process.

For the coefficient $\dot{N}_{0}m_{e}c/4D_{0}\mu$, we set several parameters in this complicated coefficient as free parameters.
Eq (1) is written as
\begin{eqnarray}
N_{0}(\gamma, \gamma_{0})&=&
\begin{cases}
 {N}_1{\gamma}^{\alpha_{1}}, &\text{$\gamma_{min}\leq\gamma\leq\gamma_{0}$},\\
 {N}_2{\gamma}^{\alpha_{2}}, &\text{$\gamma_{0}<\gamma<\gamma_{max}$}.
\end{cases}
\label{Eq:4}
\end{eqnarray}
Where $\dot{N}_{0}m_{e}c/4D_{0}\mu\gamma_{0}^{\alpha_{1}}={N}_1$, $\dot{N}_{0}m_{e}c/4D_{0}\mu\gamma_{0}^{\alpha_{2}}={N}_2$.
Suppose that appropriate boundary conditions exist at the interface between the shock region and the downstream region, the
accelerated particles $N_{0}(\gamma, \gamma_{0})$ subsequently move from the shock front to the downstream region and loses energy \citep{{1994plas.conf.....K}}.
Lorentz factor $\gamma$ can be used to evolve the particle distribution of the injected particle spectrum near the shock through the equation \citep{2018MNRAS.478.3855Z}
\begin{eqnarray}
\frac{\partial N(\gamma,t)}{\partial t}=\frac{\partial}{\partial\gamma}(\frac{d\gamma}{dt})N(\gamma,t)-\frac{N(\gamma,t)}{t_{esc}}+Q(\gamma).\;
\label{Eq:5}
\end{eqnarray}
Where the term $d\gamma/dt$ refers to synchrotron and inverse Compton cooling of the particles at time t, which includes synchrotron loss and inverse Compton loss rate corrected by Klein-Nishina (KN) effects \citep{{2005MNRAS.363..954M,2015ApJ...809...85F}}. Particles are supposed to escape from emission region with an energy independent rate of ${t_{esc}}^{-1}$ \citep{1998A&A...333..452K}, and the term $Q(\gamma)$ describes the injection rate of particles into the accelerating zone by the surrounding plasma. We set $Q(\gamma)\sim\kappa{N_{0}(\gamma, \gamma_{0})}$, and $\kappa$ is the appropriate boundary condition at the shock interface, with the unit  {$s^{-1}$}. Energy loss and escape can balance the injection of particles, which can be achieved at steady-state using the solution \citep{2009herb.book.....D}
\begin{eqnarray}
N(\gamma)=\frac{1}{c_{0}\gamma^{2}}\int_{\gamma}^{\gamma^{2}}d\gamma^{'}Q(\gamma^{'})
exp(-\int_{\gamma}^{\gamma^{'}}\frac{d\gamma^{''}}{c_{0}t_{esc}\gamma^{''}}).
\label{Eq:6}
\end{eqnarray}
 Here, we define $\gamma_{c}=1/{c_{0}}t_{esc}$ with the constant $c_{0}$, where $\gamma_{c}$ is the cooling electron break Lorentz factor \citep{2018MNRAS.478.3855Z}.
There are two regimes, fast cooling and slow cooling for the distribution of particles in the downstream flow
(e.g., \citealt{2013ApJ...763..134F,2018MNRAS.478.3855Z}). In this scenario, we define the electron energy  distribution by the  $\gamma_{min}$, $\gamma_{0}$, $\gamma_{c}$ and $\gamma_{max}$, and discuss from the following four cases:
\begin{figure}
\centering
 \includegraphics[angle=0,width=8.4cm]{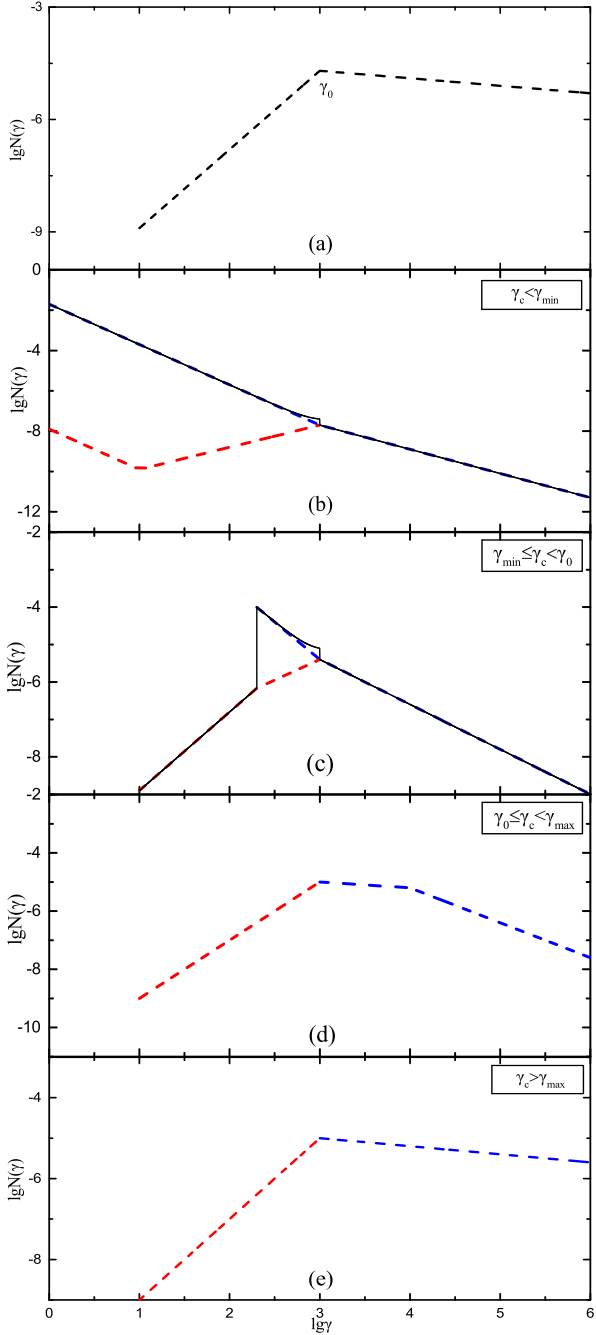}
    \caption{The electron energy distribution of the low energy particle injection (top panel) and the electron energy distribution in four different cases. For Figure (b), the red and the blue dashed line represent the particle distribution that $\gamma_{c}<\gamma<\gamma_{0}$ and  $\gamma_{c}<\gamma<\gamma_{max}$, respectively. For Figure (c), the red and the blue dashed line represent $\gamma_{min}<\gamma<\gamma_{0}$ and $\gamma_{c}<\gamma<\gamma_{max}$, respectively. For Figure (d), the red and the blue dashed line represent the three-segment spectrum of Eq (9), while for Figure (e), they represent the two-segment spectrum of Eq (10). }
 \label{sub:fig1}
\end{figure}

When the $\gamma_{c}$ satisfies the relationship $\gamma_{c}<\gamma_{min}$, the fast cooling regime of particles exists in the downstream region of the shock wave. Thus, we obtain the three-segment electron energy distribution, which is illustrated by the solid black line in Figure 1 (b). We can approximate the electron energy distribution as
\begin{eqnarray}
{N\left( {\gamma ,{\gamma _0}} \right) \approx \left\{ \begin{array}{l}
\frac{{\kappa {N_1}\gamma _{min}^{(1 + {\alpha _1})}}}{{{c_0}}}{\gamma ^{ - 2}} + \frac{{\kappa {N_2}\gamma _0^{(1 + {\alpha _2})}}}{{{c_0}}}{\gamma ^{ - 2}}\;\\
\;\;\;\;\;\;\;\;\;\;\;\;\;\;\;\;\;\;\;\;\;\;\;\;{\gamma _c} \le \gamma  \le {\gamma _{min}},\;\;\\
\frac{{\kappa {N_1}}}{{{c_0}}}{\gamma ^{\left( {{\alpha _1} - 1} \right)}} + \frac{{\kappa {N_2}\gamma _0^{(1 + {\alpha _2})}}}{{{c_0}}}{\gamma ^{ - 2}}\\
\;\;\;\;\;\;\;\;\;\;\;\;\;\;\;\;\;\;\;\;\;\;\;\;{\gamma _{min}} < \gamma  \le {\gamma _0},\;\\
\frac{{\kappa {N_2}}}{{{c_0}}}{\gamma ^{\left( {{\alpha _2} - 1} \right)}}\;\;\;\;\;\;\;{\gamma _0} < \gamma  \ll {\gamma _{max}}.
\end{array} \right.}\
\label{Eq:7}
\end{eqnarray}

When the $\gamma_{c}$ satisfies the relationship $\gamma_{min}\le\gamma_{c}<\gamma_{0}$, both slow cooling and fast cooling regime exist in the dissipative zone downstream of the shock wave. We obtain the three-segment electron energy distribution in situations such as this, which is illustrated by the solid black line in Figure 1 (c). We can approximate the electron energy distribution as
\begin{equation}
N\left( \gamma,\gamma_{0}  \right) \approx \left\{ \begin{array}{l}
\kappa N_{1}t_{esc}\gamma ^{\alpha_{1}}\;\;\;\;\;\;\;\gamma_{min}\le\gamma\le\gamma_{c},\\
\frac{\kappa N_{1}}{{c_{0}}}\gamma ^{ \left(\alpha_{1}-1\right)}+
\frac{\kappa N_{2}\gamma_{0}^{(1+\alpha_{2})}}{c_{0}}\gamma ^{-2}\\\;\;\;\;\;\;\;\;\;\; \;\;\;\;\; \;\;\;\;\; \;\;\;\;\; \;
\gamma_{c} < \gamma \le \gamma_{0},\\
\frac{\kappa N_{2}}{{c_{0}}}\gamma^{ \left(\alpha_{2}-1\right)}\;\;\;\;\;\;\;\;\gamma_{0}<\gamma \ll \gamma_{max}.
\end{array}\right.\;
\label{Eq:8}
\end{equation}

When the $\gamma_{c}$ satisfies the relationship $\gamma_{0}\le\gamma_{c}<\gamma_{max}$, the slow cooling regime of particles exists in the downstream region of the shock wave, and we obtain the three-segment spectrum in Figure 1 (d). We can  approximate the electron energy distribution as
\begin{equation}
N\left( \gamma,\gamma_{0}  \right) \approx \left\{ \begin{array}{l}
\kappa N_{1}t_{esc}\gamma ^{\alpha_{1}}\;\;\;\;\;\;\;\gamma_{min}\le\gamma\le\gamma_{0},\\
\kappa N_{2}t_{esc}\gamma ^{\alpha_{2}}\;\;\;\;\;\;\;\gamma_{0}<\gamma\le\gamma_{c},\\
\frac{\kappa N_{2}}{{c_{0}}}\gamma ^{ \left(\alpha_{2}-1 \right)}\;\;\;\;\;\;\;\gamma_{c} < \gamma \ll \gamma_{max}.
\end{array} \right.\;
\label{Eq:9}
\end{equation}

When the $\gamma_{c}$ satisfies the relationship $\gamma_{c}>\gamma_{max}$, in the downstream of the shock wave, there are both slow cooling regime and fast cooling regime. Then two-segment electron spectrum can be obtained, which is illustrated in Figure 1 (e). We can approximate the electron energy distribution as
\begin{equation}
N\left( \gamma,\gamma_{0}  \right) \approx \left\{ \begin{array}{l}
\kappa N_{1}t_{esc}\gamma ^{\alpha_{1}}\;\;\;\;\;\;\;\;\gamma_{min}\le\gamma\le\gamma_{0},\\
\kappa N_{2}t_{esc}\gamma ^{\alpha_{2}}\;\;\;\;\;\;\;\;\gamma_{0}<\gamma\ll\gamma_{max}.
\end{array} \right.\;
\label{Eq:10}
\end{equation}
To simplify the calculation, we adopt ${K}_1=\kappa N_{1}t_{esc}$ and ${K}_2=\kappa N_{2}t_{esc}$ to calculate the electron energy spectrum in four cases, where ${K}_1$ and ${K}_2$ have a covariant relationship, ${K}_2={K}_1\times {\gamma_{0}}^{(\alpha_{1}-\alpha_{2})}$.

Low energy particles are injected upstream of the shock wave and are accelerated to form a power-law distribution. It can be seen that (a) in Figure 1 shows the electron energy distribution of the low energy particle injection, extending from the minimum energy to the maximum energy.
 We fix the values of $\gamma_{0}$, $\gamma_{min}$ and $\gamma_{max}$, change the values of $\gamma_{c}$ in four cases. In figure 1, we assume $\gamma_{0}=1\times10^3$, $\gamma_{min}=10$, $\gamma_{max}=1\times10^6$, $\alpha_{1}=2.1$, $\alpha_{2}=-0.2$. The basic physical parameters we assume are as follows: the magnetic field $B=0.37\rm~G$, the emission region size $R=1\times10^{16}\rm~cm$, the Doppler factor $\delta=30$.

\section{The theoretical photon spectrum}

Without doubt,
the SSC model has been widely used to explain the emission energy spectrum of blazars, which spans from the optical bands to the X-ray bands, and even the high-energy $\gamma$-ray bands \citep{1981ApJ...243..700K}. The one-zone SSC model postulates there exists a spherical emission region with non-thermal relativistic electrons and a magnetic field.
We postulate that the electron distribution is isotropic. Once the electron distribution in the co-moving frame is determined, we can calculate the emission coefficient due to synchrotron emission,
\begin{eqnarray}
j_{\rm syn}(\nu)&=&\frac{\sqrt{3}e^{3}B }{4\pi m_{e}c^{2}}\int N(\gamma, \gamma_{0})\nonumber\\&\times&F(\frac{4\pi m_{e}c\nu}{3eB\gamma^{2}})d\gamma~~\rm erg~cm^{-3}~s^{-1}~Hz^{-1}\;.
\label{Eq:11}
\end{eqnarray}
Here, $F(x)$ is the modified Bessel function. In the conventional sense, synchrotron emission is rifely accompanied by self-absorption. The self-absorption of synchrotron can be simply regarded as the absorption of soft photons. We can obtain the self-absorption coefficient,
\begin{eqnarray}
k_{\rm syn}(\nu)=&-\frac{\sqrt{3}e^{3}B }{8\pi m_{e}^{2}c^{2}}\int\gamma^{2}F(\frac{4\pi m_{e}c\nu}{3eB\gamma^{2}})\nonumber\\
&\times \frac{\partial}{\partial\gamma}\biggl[\frac{N(\gamma, \gamma_{0})}{\gamma^{2}}\biggr]d\gamma~~\rm cm^{-1}\;.
\label{Eq:12}
\end{eqnarray}
The emission intensity of electron synchrotron emission can be obtained by solving the emission transfer equation of spherically symmetric structure \citep{1999ApJ...514..138K},
\begin{eqnarray}
I_{\rm syn}(\nu)=&\frac{j_{\rm syn}(\nu)}{k_{\rm syn}(\nu)}\biggl[1-{\frac{2}{\tau^{2}}(1-{\tau{e^{-\tau}}}-e^{-\tau})}\biggr]~~\nonumber\\&\rm erg~cm^{-2}~s^{-1}~Hz^{-1}\;,
\label{Eq:13}
\end{eqnarray}
where $\tau=2Rk(\nu)$. Assuming that the intensity of the synchrotron is uniform throughout the emission region, the uniform emission intensity within the emission area is usually expressed as the emission intensity at the center of the blob multiplied by 3/4  in order to correct for the impact of the synchrotron emission intensity decreasing with the increase of the blob radius \citep{{1979A&A....76..306G,1999ApJ...514..138K}}. Hence, the soft photon number density of synchrotron radiation is,
\begin{equation}
n(\epsilon_{\rm syn})={\frac{3}{4}}\frac{4\pi}{hc\epsilon_{\rm syn}}\frac{j_{syn}(\nu)}{k_{syn}(\nu)}[1-e^{-k_{syn}(\nu)R}]~~\rm cm^{-3}\;,
\label{Eq:14}
\end{equation}
the differential photon generation rate of the inverse Compton scattering is,
\begin{eqnarray}
q(\epsilon_{\rm ic})=&\int n(\epsilon_{\rm syn})d\epsilon_{\rm syn}\times\int N(\gamma, \gamma_{0})\nonumber\\&\times C(\epsilon_{\rm ic}, \gamma, \epsilon_{\rm syn})d\gamma~~\rm cm^{-3}~s^{-1}\;.
\label{Eq:15}
\end{eqnarray}
Where $C(\epsilon_{\rm ic}, \gamma, \epsilon_{\rm syn})$ is the Compton function factor \citep{jones1968calculated}.
The emission coefficient of inverse Compton scattering is deduced from the photon production rate,
\begin{equation}
j_{\rm ic}(\nu)=\frac{h}{4\pi}\epsilon_{\rm ic}q(\epsilon_{\rm ic})~~\rm erg~cm^{-3}~s^{-1}~Hz^{-1}\;.
\label{Eq:16}
\end{equation}
Here, $\epsilon$ is the photon energy being scattered, and it's equal to $h\nu/m_{e}c^{2}$ \citep{2019ApJ...873....7Z}. After that, by restricting the range of differential photon generation rates, we can calculate the intensity of synchronous self-compton emission,
\begin{equation}
I_{\rm ic}(\nu)=j_{\rm ic}(\nu)R~~\rm erg~cm^{-2}~s^{-1}~Hz^{-1}\;.
\label{Eq:17}
\end{equation}
It is a truism that the source generated high-energy photons may be absorbed by the extragalactic background light (EBL) due to the photon-photon pair generation process \citep{{2005ApJ...618..657D,2015ApJ...814...20F}}, hence we consider the absorption effect of EBL.
Using the Doppler beaming effect to modify the radiation intensity, we obtain the emission flow in the laboratory coordinate system,
\begin{equation}
 F_{obs}(\nu)={\frac{\pi R^{2}\delta^{3}(1+z)}{{D_{L}}^{2}}}\biggl[I_{\rm syn}(\nu)+I_{\rm ic}(\nu)\biggr]\times e^{-\tau(\nu,z)},
 \label{Eq:18}
\end{equation}
where $D_{L}$ is the luminosity distance, $z$ is the cosmological red shift. $\delta={[\Gamma(1-\beta\cos\theta)]}^{-1}$ is the Doppler factor, where $\Gamma$ is the Lorentz factor, $\theta$ is the Angle between the velocity of the blob vector and the line of sight.
The interaction between electrons and photons results in the absorption depth of the VHE photons, which is represented here in terms of $\tau(\nu,z)$ \citep{{2005ApJ...618..657D,2004A&A...413..807K}}.
\section{Application to Mrk 421}
Mrk 421 is the source of the BL Lac object with redshift at z=0.031, and it has invisible emission or absorption lines. It is the first BL Lac object detected by the Energetic Gamma Ray Experiment Telescope \citep{1992ApJ...401L..61L} at energies greater than 100 MeV which is a typical  high-synchrotron-peaked (HSP) source. Furthermore, its SED has been extensively studied and can be represented by two peaked components with distinct characteristics \citep{{1992Natur.358..477P,1995PASP..107..803U,1997ARA&A..35..445U}}. In addition, the SED can be well fitted by a one-zone SSC model \citep{{2013PASJ...65..109C,2016ApJ...819..156B}}. The observations from the radio band to the soft X-ray band indicate that this portion of the SED is caused by the distribution of relativistic electrons from the radiation during the synchrotron process. The radiation we obtain from the gamma-ray band is most likely due to the inverse Compton scatter caused by the high energy electrons in the synchrotron radiation, as evidenced by the simultaneous correlation of the low energy and the high energy SED components \citep{{2007A&A...462...29G,2015A&A...576A.126A}}.
Although the multi-band energy spectrum of Mrk 421 has been extensively researched for a long time, the nature of the jets, the location of gamma-ray emission and the emission mechanism are still not understood.

Since we have determined the electron spectrum under the assumption of Eq (9), we use it to calculate the SED in the homogeneous SSC model.
We postulate that a spherical blob located downstream is filled with uniform magnetic fields and ultra-relativistic electrons.
To simplify the operation, we first rewrite the electron spectrum, the density normalization coefficient ${K}_1=\kappa N_{1}t_{esc}$ is defined, and ${K}_2={K}_1\times {\gamma_{0}}^{(\alpha_{1}-\alpha_{2})}$. Therefore, Eq (9) can be rewritten as follows.

When the $\gamma_{c}$ satisfies the relationship $\gamma_{0}\le\gamma_{c}<\gamma_{max}$,  the electron energy distribution can be written as
\begin{equation}
N\left( \gamma,\gamma_{0}  \right) \approx \left\{ \begin{array}{l}
{K}_1\gamma ^{\alpha_{1}}\;\;\;\;\;\;\;\;\;\;\;\;\gamma_{min}\le\gamma\le\gamma_{0}\\
{K}_2\gamma ^{\alpha_{2}}\;\;\;\;\;\;\;\;\;\;\;\;\gamma_{0}<\gamma\le\gamma_{c}\\
{K}_2\gamma_{c}\gamma ^{  \left( {\alpha_{2} - 1}\right)}\; \;\gamma_{c} < \gamma \ll \gamma_{max}
\end{array} \right.\;.
\label{Eq:19}
\end{equation}
\begin{figure}
\centering
 \includegraphics[angle=0,width=8cm]{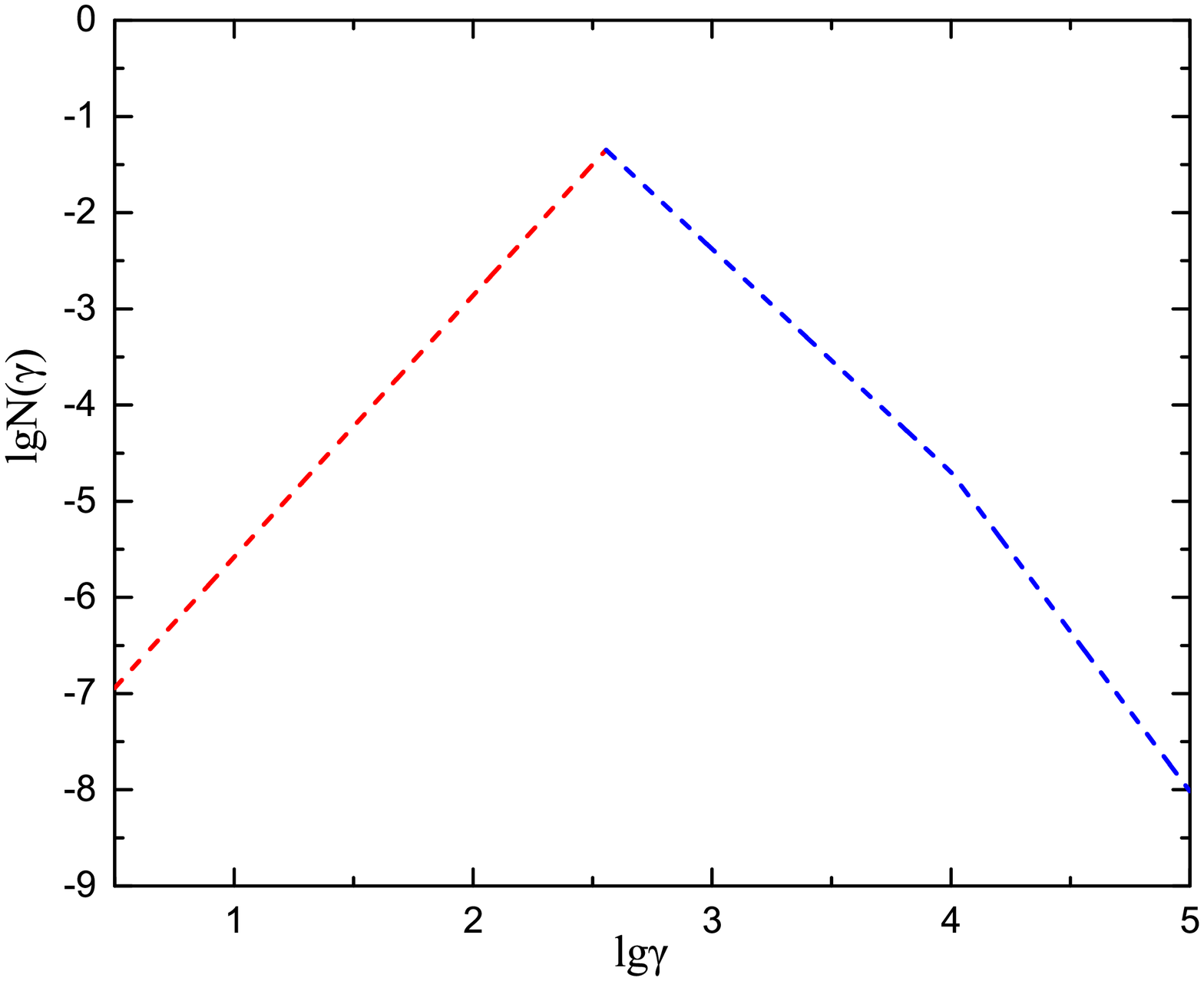}
 \includegraphics[angle=0,width=8cm]{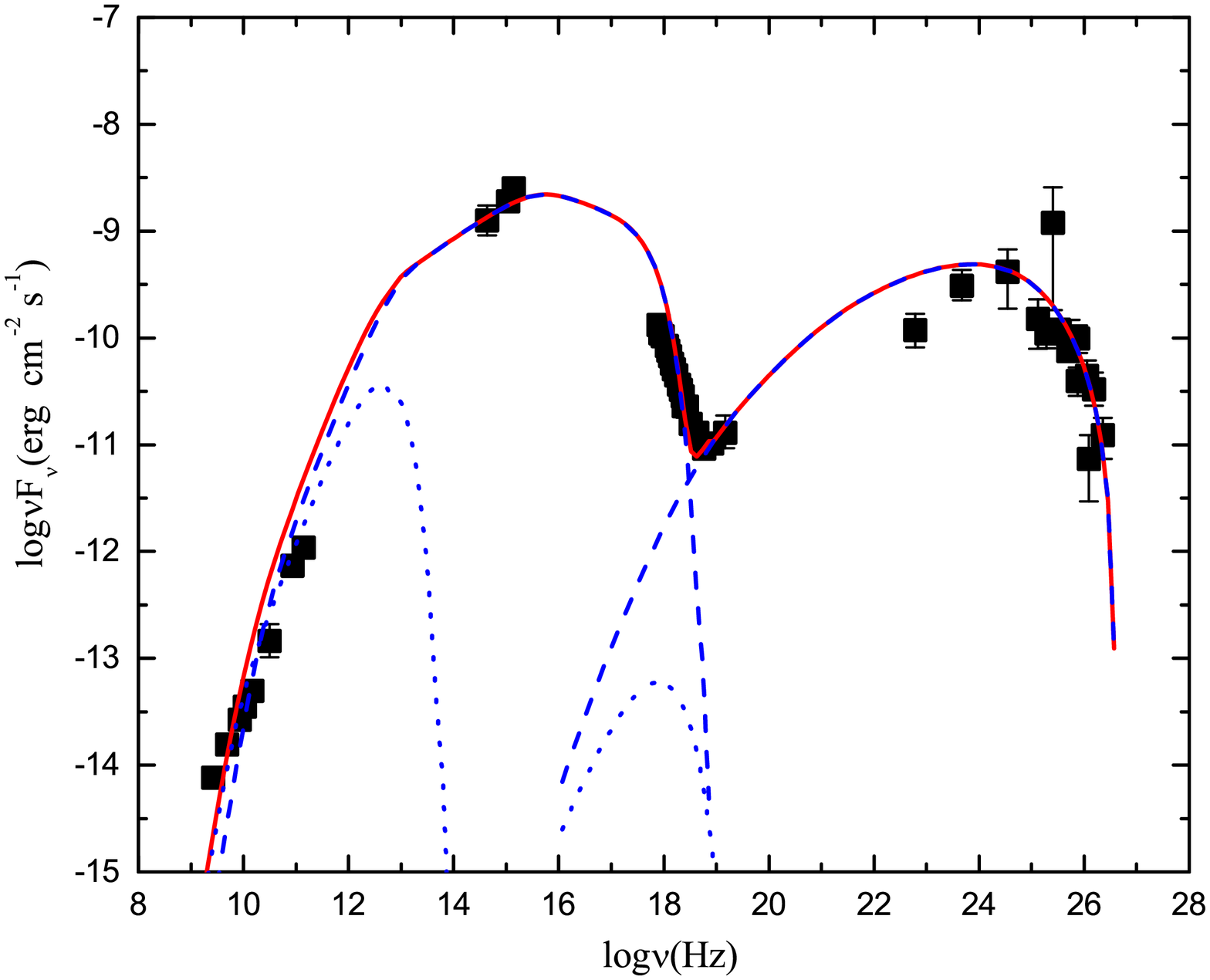}
    \caption{The electron energy spectrum (top panel) and the comparison of the predicted multi-wavelength spectrum with the observed data of Mrk 421 (bottom panel). For the electron energy spectrum, the dotted line shows the three-segment spectrum, corresponding to Eq (19).
    For the figure at the bottom, the dotted blue line represents  $\gamma\leq\gamma_{0}$, the dashed blue line represents $\gamma>\gamma_{0}$, and the solid red line represents the total spectrum of the superpositions. In this figure, the data points come from \cite{2017ApJ...842..129C} and references therein. The parameters selected in the two pictures are the same, as shown in Table 1.}
 \label{sub:fig3}
\end{figure}
We presume that the relativistic electrons are in a stable state during the observation period. Using the derived electron energy distribution in the downstream flow from the plane shock, we can calculate the SED within the framework of a homogeneous model on the basis of the postulated model results.
In order to investigate whether the SED can interpret multi-wavelength spectra, we apply our model to BL Lac object Mrk 421.

Figure 2 shows our predicted energy spectrum distribution from the radio band to the TeV $\gamma$-ray band. The  dotted  blue line represents $\gamma\leq\gamma_{0}$, the dashed blue line represents $\gamma>\gamma_{0}$, and the solid red line represents the total spectrum of the superpositions. Meanwhile, we also show the corresponding electron spectrum of the SED fitting, in order that we can analyze it more intuitively. We change the values of the model parameters  $\gamma_{min}$, $\gamma_{0}$, $\gamma_{c}$, $\gamma_{max}$, $\alpha_{1}$, $\alpha_{2}$, $\delta$, $B$ and $R$ until we get a reasonable qualitative fitting of multi-wavelength spectral data.
As can be seen from the figure: 1) For the  electron energy spectrum of $\gamma_{0}<\gamma\leq\gamma_{c}$,  the escape process dominates the energy losses, resulting in a flat energy spectrum. Above the cooling break energy $\gamma_{c}$, the cooling process dominates the energy losses, leading to steepening of the spectrum;
2) Under reasonable hypotheses of parameters, the model we utilized was able to roughly reproduce the observed spectra;
3) the part of $\gamma>\gamma_{0}$ in the electron spectrum contributes more to the radiation energy spectrum, while the part of $\gamma\leq\gamma_{0}$ contributes less to the radiation energy spectrum.

\begin{table}[htb]\footnotesize
\caption{The physical model parameters for Mrk 421}
\begin{tabular}{@{}ll@{}}
\hline\hline
parameters & double broken power law  \\
\hline
Doppler factor       &$\delta=30$\\
Magnetic field       &$B=0.37$~G\\
Emission region size &$R=1\times10^{16}$~cm\\
\hline
Minimum electron Lorentz factor       &$\gamma_{min}=1$\\
Cooling  break Lorentz factor    &$\gamma_{c}=1\times10^{4}$\\
Injected  electron Lorentz factor   &$\gamma_{0}=3.6\times10^{2}$ \\
Maximum electron Lorentz factor       &$\gamma_{max}=1.12\times10^{5}$\\
Index of the electron   &$\alpha_{1}=2.72,\alpha_{2}=-2.32$ \\
Reduced chi-square               &$\chi^{2}=0.1409$ \\
\hline
\end{tabular}
\end{table}
In the model, we can estimate the jet parameters when provided with a broadband SED, the parameters are shown in Table 1. $\gamma_{0}$ as the injected Lorentz factor of our assumption of low energy particles, its low value from the obtained results is consistent with our hypothesis.  The reduced chi-square value obtained by the fitting is approximate to 0, indicating that the fitting results of the model are reasonable. In this scenario, we can obtain the magnetic field energy density through the relationship  $U_{B}=B^{2}/8\pi=0.0054 \rm~ erg\rm~ cm^{-3}$ and to obtain the electron energy density by calculating $U_{e}=m_{e}c^{2}\int{N(\gamma)}{\gamma}d\gamma$ \citep{{1993MNRAS.264..228C,2018MNRAS.478.3855Z}}, as well as the proton energy density $U_{p}=m_{e}c^{2}\int{N(\gamma)}d\gamma$. By this means, the value of the jet power can also be calculated by
\begin{equation}
  P_{j}=P_{e}+P_{B}+P_{p}=\pi R^{2}c\delta^{2}(U_{e}+U_{B}+U_{p})\\
  \label{20}
\end{equation}
 \citep{{celotti2008the,2017MNRAS.464..599D}}. Thus, we can calculate the jet power as $9.3\times10^{43}\rm ~erg \rm~ s^{-1}$. Such a value for the jet power is commonly found in the jet of BL Lac object Mrk 421 (e.g., \citealt{2017MNRAS.464..599D,2017ApJ...842..129C}). For other parameters, the Doppler factor of our fits are consistent with the Doppler factor range $\delta\gtrsim 30$, found for the SSC fit to the same souce Mrk 421 by \cite{{2008ApJ...686..181F}} which detected by HEGRA and RXTE in March 2001. As is known to all, the AGN emission radius of the blob is generally smaller than other galaxies, approximately less than 0.1 pc. To some extent, the model parameters determined are expected to be rational.

Thanks to the particle spectrum Eq (1) contains some crucial parameters of physical significance, we utilize the obtained physical parameters to reverse them. On the basis of some formulas given by \cite{{2019ApJ...873....7Z}}, we find that some other parameters can be deduced by using the model parameters, for instance, we substitute the power law indices $\alpha_{1}$ and $\alpha_{2}$ into the characteristic polynomials Eq (2) and Eq (3) to invert the dimensionless parameters $\hat{A}$ and $\hat{C}$.
Therefore, we can determine that $\hat{A}=-1.6$ and $\hat{C}=6.31$. Due to $\mu=0.25[(2+\hat{A})^{2}+4\hat{C}]^{1/2}$, we
substitute the values of $\hat{A}$ and $\hat{C}$ into $\mu$, then obtain the result $\mu=1.26$. We set $K_{1}=\chi N_{1}t_{esc}$, and the typical escape timescales that we assume are as follows \citep{2010A&A...515A..18W},
\begin{equation}
t_{esc}=\frac{1.5\times{R}}{c},
\label{Eq:21}
\end{equation}
where $c$ is the speed of light. We calculate that the model parameter $K_{1}=5\times10^{-9}\rm~cm^{-3}$, and we also can get the relation for $K_{1}$,
\begin{equation}
K_{1}={\kappa}{t_{esc}}\frac{\dot{N}_{0}m_{e}c}{4D_{0}\mu\gamma_{0}^{\alpha_{1}}}.\\
\label{Eq:22}
\end{equation}
Here, $D_{0}$ is the momentum diffusion rate constant and $\dot{N}_{0}$ is the continuous injection rate of particles.
The momentum diffusion rate constant is closely related to the momentum diffusion coefficient and is one of the important parameters of the particle transport equation. Considering both Eq (21) and Eq (22), the relationship  between  $D_{0}$ and  $\dot{N}_{0}$ is as follows,
\begin{eqnarray}
\dot{N}_{0}&=&{1.65\times10^{10}}\rm~ p^{-1}~s^{-1}~cm^{-3}\nonumber\\&\times& \biggl(\frac{D_{0}}{s^{-1}}\biggr)\biggl(\frac{R}{10^{16}~\rm cm}\biggr)^{-1} \biggl(\frac{\kappa}{s^{-1}}\biggr)^{-1}\,, \
\label{Eq:23}
\end{eqnarray}
due to $\hat{A}=A_{0}/D_{0}$, $\hat{C}=C_{0}/D_{0}$ \citep{{2016ApJ...833..157K,2019ApJ...873....7Z}},  we can obtain
\begin{eqnarray}
A_{0}&=&{-9.67\times10^{-11}\rm~ s^{-1}}\nonumber\\&\times& \biggl(\frac{R}{10^{16}~\rm cm}\biggr)\biggl(\frac{\kappa}{s^{-1}}\biggr)\biggl(\frac{\dot{N}_{0}}{p^{-1}\rm~ s^{-1}\rm~ cm^{-3}}\biggr) \,,\
\label{Eq:24}
\end{eqnarray}
where $A_{0}$ is the first-order momentum gain rate constant in the unit of $s^{-1}$, we also can obtain $C_{0}$,
\begin{eqnarray}
C_{0}&=&{3.82\times10^{-10}\rm~ s^{-1}}\nonumber\\&\times& \biggl(\frac{R}{10^{16}~\rm cm}\biggr)\biggl(\frac{\kappa}{s^{-1}}\biggr)\biggl(\frac{\dot{N}_{0}}{p^{-1}\rm~ s^{-1}\rm~ cm^{-3}}\biggr) \,, \
\label{Eq:25}
\end{eqnarray}
where $C_{0}$ is a shock regulated escape rate constant.

\section{Discussion}
Various mechanisms can be used to confirm the required distribution of emitted particles \citep{{2014ApJ...780...87M,2012ApJ...745...63S,1978MNRAS.182..147B}}, for instance, the particles accelerated by the diffusion shock wave can generate a new particle distribution. Our model considers plane shock waves propagating along cylindrical jets.
Low energy particles are injected upstream of the shock wave and are accelerated to form a power-law distribution, then they drift downstream and lose energy. We obtain the particle distribution under four different conditions according to the different value of $\gamma_{c}$.
The electron above the equilibrium energy produces a power-law distribution, so does the electron below the equilibrium energy.
Assuming appropriate model parameters, we utilize the obtained power law distribution to calculate the spectrum energy distribution when $\gamma_{0}\le\gamma_{c}<\gamma_{max}$ under the SSC framework and applied it to the extreme BL Lac object Mrk 421, thus reproducing the multi-wavelength SED.
From the obtained multi-wavelength spectrum energy distribution of Mrk 421, it can be seen that the components of the electron spectrum affect the photon spectrum. In the total emission energy spectrum after the superposition, we can see that the calculated result of  $\gamma>\gamma_{0}$  is  similar to that of the superposition spectrum. The portion of the electron spectrum  $\gamma>\gamma_{0}$ contributes more to the radiation energy spectrum. On the contrary, the electron spectrum  $\gamma\leq\gamma_{0}$  contributes little to the energy of the total spectrum.
Actually, we take the value of the break energy $\gamma_{c}$ as free parameters. We find that for $\gamma\leq\gamma_{0}$, assuming the value of $\gamma_{c}$ close to $\gamma_{0}$, then the energy spectrum will contribute more to the emission spectrum. Otherwise, we can ignore the first electron spectrum. The feasible reason we speculate is due to the magnitude of the energy.

As shown in Figure 2, the fitting of our model to Mrk 421 is not very well in the radio band. Previous works using one-zone SSC models have rarely been able to interpret the radio band data perfectly, this is one of the drawback of the model. The emission in the radio band of the blazar is generally considered to be outside the main emission region. On account of the superposition of synchrotron spectra from a range of jet locations where the optical depth approximately equal to 1, and cannot be formed by the same component as the low-frequency peak emission process because the radio band must be self-absorbed \citep{2018ApJ...853....6L}.

In general, the shock wave physics we investigated is constrained by the minimum energy of the shock wave accelerating particle, which is very model-dependent. The maximum energy a particle can achieve in shock acceleration is closely related to the life of the shock wave. If the life of the shock wave is limited, then the acceleration rate of the shock wave is an important factor. In the previous section we deduced the relation between the dimensionless momentum diffusion rate constant and the continuous injection rate of particles $\dot{N}_{0}$.  From the process of deriving the photon spectrum, we find that $\dot{N}_{0}$ has a great influence on the emission flux density, and the increase of the continuous injection rate will lead to the upward shift of the energy spectrum distribution. \cite{2006ApJ...647..539B} use the $\dot{N}_{0}/D_{0}$  relation as the coefficient when deriving the steady state green's function. We note that the MHD acceleration timescale must exceed the gyroscopic period of the accelerating electron, this fact can be used to quantify the momentum diffusion rate constant refer to \cite{2019ApJ...873....7Z}.

The particle injection mechanism is rarely known, nonetheless, it is still possible to perform some calculations on the parameters related to the injection mechanism.
According to some relations in the particle injection equation, we can use the obtained parameters to derive some parameters such as the momentum diffusion rate of the particle and the continuous injection rate of  particles. Previous efforts assumed the continuous injection rate of particles in the model \citep{2019ApJ...873....7Z}, and  the value of the continuous injection rate is approximately $1\times10^{20}$ with the unit $p^{-1}\rm~ s^{-1}\rm~ cm^{-3}$. Similarly, the relation between the shock  acceleration rate constant and the  momentum diffusion rate constant can also be deduced. It can be perceived from the relation we determine the value of $\dot{N}_{0}$ is approximately $1.65\times10^{10}$,  mainly because we cannot obtain the value of $D_{0}$ at this time, so we ignore its magnitude.
Considering first-order Fermi acceleration and magnetic field energy acceleration, shock acceleration dominates the process if the first-order momentum gain rate constant $A_{0}>0$ \citep{2018ApJ...853....6L}. When we take into account the adiabatic expansion energy loss, the adiabatic loss is dominant, where $A_{0}$ can be less than 0. This also leaves us with the possibility to investigate more deeply the relationship between the parameters in future work, on account of the complexity of the parameters of Eq (1).

\acknowledgments
We thank the anonymous referee for very constructive and helpful comments and suggestions, which greatly helped us to improve our paper. This work was partially supported by the National Natural Science Foundation of China (Grant Nos.11763005 and 11873043), and the Science and Technology Foundation of Guizhou Province (QKHJC[2019]1290).

\bibliographystyle{spr-mp-nameyear-cnd}
\bibliography{article}

\end{document}